\documentclass{article}

\usepackage{arxiv}

\usepackage[utf8]{inputenc} 
\usepackage[T1]{fontenc}    
\usepackage{hyperref}       
\usepackage{url}            
\usepackage{booktabs}       
\usepackage{amsfonts}       
\usepackage{nicefrac}       
\usepackage{microtype}      
\usepackage{doi}

\usepackage{color}
\usepackage{bm}
\usepackage[title]{appendix}
\usepackage{float}
\usepackage{graphicx}
\usepackage{amssymb}
\usepackage{amsmath}
\usepackage{graphicx}
\usepackage{enumerate}
\usepackage{blindtext}
\usepackage{booktabs}
\usepackage{multirow}
\usepackage[backend=bibtex, style=authoryear, maxbibnames = 40, natbib]{biblatex}
\title{Bin-Conditional Conformal Prediction of Fatalities from Armed Conflict\thanks{The research was funded by the European Research Council, project H2020-ERC-2015-AdG 694640 (ViEWS) and Riksbankens Jubileumsfond, grant M21-0002 (Societies at Risk),  the European Research Council under  Horizon Europe (Grant agreement No. 101055176, ANTICIPATE),  Norwegian Research Council (NFR) grant 334977 (The Uncertainty of Forecasting Fatalities in Armed Conflict (UFFAC)) and the Center for Advanced Study (CAS) at the Norwegian Academy of Science and Letters.}}

\author{David Randahl \\
	Dep. of Peace and Conflict Research \\
        Uppsala University\\
	\texttt{david.randahl@pcr.uu.se} \\
	\And
	Jonathan P Williams \\
	Dep. of Statistics\\
	North Carolina State University\\
	\texttt{jwilli27@ncsu.edu} \\
	 \AND
	 Håvard Hegre \\
  Peace Research Institute Oslo \\
	Dep. of Peace and Conflict Research\\
	Uppsala University\\
	\texttt{hhegre@prio.org}
}

\renewcommand{\shorttitle}{Bin-Conditional Conformal Prediction of Fatalities from Armed Conflict}

\begin{document}

\maketitle

\begin{abstract}

Forecasting armed conflicts is a critical area of research with the potential to save lives and mitigate suffering. While existing forecasting models offer valuable point predictions, they often lack individual-level uncertainty estimates, limiting their usefulness for decision-making. Several approaches exist to estimate uncertainty, such as parametric and Bayesian prediction intervals, bootstrapping, quantile regression, but these methods often rely on restrictive assumptions, struggle to provide well-calibrated intervals across the full range of outcomes, or are computationally intensive. Conformal prediction offers a model-agnostic alternative that guarantees a user-specified level of coverage but typically provides only marginal coverage, potentially resulting in non-uniform coverage across different regions of the outcome space. In this paper, we introduce a novel extension called bin-conditional conformal prediction (BCCP), which enhances standard conformal prediction by ensuring consistent coverage rates across user-defined subsets (bins) of the outcome variable. We apply BCCP to simulated data as well as the forecasting of fatalities from armed conflicts, and demonstrate that it provides well-calibrated uncertainty estimates across various ranges of the outcome. Compared to standard conformal prediction, BCCP offers improved local coverage, though this comes at the cost of slightly wider prediction intervals.
\end{abstract}

\keywords{Conformal prediction \and Conflict forecasting \and Prediction intervals \and Skewed distributions}

\section{Introduction}

In recent years, the forecasting of armed conflict has become a central focus in conflict research, with numerous projects aiming to predict a wide range of political and conflict-related outcomes \citep[e.g.][]{hegre2019views,hegre2021views2020,mueller2018reading,bell2016coup,morgan2019varieties,vesco2022united,williams2024}. While significant progress has been made, most forecasts still rely primarily on point predictions without accompanying measures of uncertainty at the individual prediction level. Instead, model performance is typically evaluated through aggregate error metrics, such as mean squared error or accuracy rates.

This reliance on point predictions alone is problematic for several reasons. First, point predictions, while often informative, are inherently uncertain. In many cases, they will be incorrect, and without uncertainty estimates, their utility to stakeholders is limited. This limitation is particularly concerning in fields where outcomes are rare but have potentially high impacts, such as armed conflict forecasting. In such cases, predicted outcomes often cluster around zero due to the rarity of high-fatality events. This can lead to an underestimation of extreme events unless uncertainty is explicitly accounted for \citep{hegre20242023}.

Second, effective communication of forecast results requires uncertainty quantification. Stakeholders, such as policymakers, are often more interested in the range of possible outcomes than in a single point estimate. For example, a best-guess point prediction of zero fatalities in a given month for a particular country may mask a non-negligible probability of severe conflict. The best point prediction for a non-violent but stable country in a given month is typically 0 fatalities, but the probabilities of more than 1,000 deaths or even 100,000 may be alarmingly high. Performing a simple prediction of the probability of exceeding a fatality threshold \citep[as in ][]{hegre2021views2020} can partially address this, but the resulting probabilities are not necessarily well-calibrated and do not provide the full range of the associated uncertainty. Well-calibrated prediction intervals can therefore help to communicate risk effectively and to prevent the misinterpretation of single-valued forecasts. Similarly for binary outcomes, providing prediction intervals for the probability of an event can help to communicate the uncertainty in the models' predictions.

Third, aggregate performance metrics, such as the area under the receiver operating characteristic curve (AUC-ROC) or continuous rank probability scores, provide useful information about model performance but do not indicate the level of uncertainty for individual predictions. A model may perform well overall but still fail to quantify uncertainty correctly for different subpopulations or data regions.

A variety of methods exist to estimate uncertainty in predictive modeling.  Frequentist and Bayesian parametric approaches allow for uncertainty estimation via the sampling distribution of a pivot statistic or a posterior distribution, respectively.  Frequentist prediction sets, however, require correct specification of the distribution of the data to achieve finite-sample coverage of prediction sets at any specified level, and otherwise rely on asymptotic justifications.  Bayesian credible prediction sets are not naturally calibrated to have any repeated sampling properties (i.e., they are not confidence sets), but asymptotically they may exhibit coverage at their corresponding frequentist-specified level.  Resampling techniques such as bootstrapping and jackknifing provide another alternative, but they can be computationally expensive and may not always offer well-calibrated uncertainty estimates. Similarly, quantile regression can also provide prediction intervals, but these are model-specific and may also not be well-calibrated in out-of-sample predictions \citep[for reviews of techniques for obtaining prediction intervals, see for instance][]{tian2022methods, hesterberg2015teachers, zhang2020random}.

Conformal prediction offers a model-agnostic alternative for creating prediction intervals which under certain conditions guarantees a user-specified level of coverage, i.e., that a user-specified minimum proportion of the future outcomes actually fall within the prediction intervals. Yet, standard conformal prediction (SCP) algorithms only maintain marginal coverage and do not guarantee uniform levels of local coverage across different subsets of the data.

In this paper, we introduce a novel approach called \textit{bin-conditional conformal prediction} (BCCP) to address these limitations of the SCP. BCCP extends SCP by ensuring that prediction intervals achieve appropriate coverage rates not only in the aggregate but also within user-defined subsets (bins) of the data. This method provides a flexible, non-parametric framework for uncertainty quantification that does not require restrictive model assumptions, while improving local coverage calibration.  The BCCP approach is a hybrid of the label-conditional, or \textit{Mondrian}, conformal prediction approach, and the SCP approach for real-valued outcomes. This method is compatible with both inductive and transductive routines, though we focus on the inductive version here.  Finite-sample, label-conditional validity of Mondrian conformal prediction is established in \cite{vovk2005algorithmic}.  Label-conditional validity is distinct from notions of covariate/feature/object-conditional validity, e.g., as in \cite{chernozhukov2021distributional} and \cite{ sesia2021conformal}.

We demonstrate the advantages of BCCP using both simulated data and real-world data from the Violence Early Warning System (ViEWS) fatality forecasting model. Our results show that BCCP consistently achieves the desired coverage levels across different subsets of the data, while SCP tends to over- or under-cover in certain regions. We also compare BCCP to alternative methods for obtaining prediction intervals. Here too we find that alternative methods often fail to provide well-calibrated uncertainty estimates across all ranges of the prediction target compared to the BCCP, which maintains the desired coverage rates across the user-specified ranges (bins).

While this paper focuses on forecasting fatalities from armed conflict, the BCCP method is applicable to any domain requiring uncertainty quantification for individual predictions. It can be used regardless of many distributional characteristics of the outcome, including for estimating uncertainty in predicted probabilities for binary or multinomial outcomes. We conclude by discussing potential future directions, including extensions of BCCP to other data structures and further evaluations against alternative uncertainty estimation techniques. The method is implemented in the R package \texttt{pintervals} and is available on CRAN.

\section{Conformal Prediction}
Conformal prediction is a general purpose, model-agnostic, method that, given any point prediction, $\hat{y}$, any desired error rate, $\alpha\in [0,1]$, and any measure of dissimilarity (non-conformity), can be used to create a prediction interval around $\hat{y}$ that with probability of $1-\alpha$ contains the true value/label $y$. Under the condition that the dissimilarity scores are \textit{exchangeable} (e.g., independent and identically distributed), conformal prediction sets are mathematically guaranteed to be calibrated to their user-specified level of coverage in repeated sampling \citep{vovk2005algorithmic,shafer2008tutorial}.

Conformal prediction is especially attractive because it is model-agnostic, i.e., that it can be applied to any prediction algorithm that outputs a point prediction, regardless of whether the point prediction is regression or classification and whether or not the prediction algorithm outputs any uncertainty measures. This means that conformal prediction can be applied on top of machine learning algorithms such as random forests, boosted tree models, and support vector machines without requiring any further adaptation \citep{fontana2023conformal,shafer2008tutorial}. Since its introduction in 1999 \citep{saunders1999transduction,vovk1999machine}, conformal prediction has been applied to a wide range of prediction problems and algorithms \citep[for a review, see][]{fontana2023conformal}.

\subsection{Assumptions of Conformal Prediction}
Conformal prediction applies to settings for which we wish to obtain a $1-\alpha$ level prediction interval for some label $y_{n+1}$, associated with feature $x_{n+1}$, based on a set of $n$ example feature-label pairs, $(x_1,y_1), \dots, (x_n,y_n)$.  The conformal prediction set at level $1-\alpha \in [0,1]$ associated with $x_{n+1}$ will include the true label $y_{n+1}$ with probability at least $1-\alpha$, provided the non-conformity scores for $(x_1,y_1), \dots, (x_n,y_n), (x_{n+1},y_{n+1})$ are exchangeable.  The non-conformity score for $(x_i,y_i)$ is a measure of dissimilarity between $(x_i,y_i)$ and $(x_1,y_1), \dots,(x_{i-1},y_{i-1})$, $(x_{i+1},y_{i+1}), \dots, (x_{n+1},y_{n+1})$.  A non-conformity score can be constructed from any real-valued function that is invariant to the order of points $(x_1,y_1), \dots,(x_{i-1},y_{i-1}), (x_{i+1},y_{i+1}), \dots, (x_{n+1},y_{n+1})$, but is generally taken as any real-valued error function for a predictive algorithm such as: the absolute or squared error between the predicted and observed values for regression problems, distance to nearest neighbor in nearest neighbor classification, or one minus the soft-max activation function components of a neural network \citep[e.g.,][]{dey2023conformal}, etc.

To obtain the $1-\alpha$ prediction set for $y_{n+1}$ we calculate the non-conformity score for our test case using $x_{n+1}$ and a range of hypothetical values which $y_{n+1}$ may have. If the support of the prediction values is finite (as in classification), then all possible labels are considered as the hypothetical values, while if the support is infinite (as in real-valued regression), then a grid of reasonable hypothetical values based on $x_{n+1}$ is iterated over. For each hypothetical $y_{n+1}$ value a \textit{p-value} is computed as the proportion of non-conformity scores in the training data that are as large or larger than the non-conformity score for the hypothetical $y_{n+1}$ value. The $1-\alpha$ prediction set is constructed by including in it all hypothetical values of $y_{n+1}$ corresponding to a p-value larger than $\alpha$ \citep{shafer2008tutorial, toccaceli2019combination}.\footnote{In regression problems, the non-conformity measure is usually a prediction error function. In these cases, the non-conformity score will increase as the value moves away from $\hat{y}$, and thus it is sufficient to find the hypothetical lower and upper edge cases which produce p-values that are less than $\alpha$.}

\subsection{Transductive and Inductive Conformal Prediction}
The non-conformity scores can be obtained either transductively or inductively. The main difference between these approaches is that transductive conformal prediction requires the prediction algorithm to be re-trained for every new test case, while inductive conformal prediction splits the training data into a training set and a calibration set. In the inductive case, the the prediction algorithm is then trained on the training set, and out-of-sample predictions and associated non-conformity scores are calculated using the calibration partition. These non-conformity scores are then used when creating the prediction set for the new test case \citep{toccaceli2019combination, papadopoulos2008inductive}. For computationally demanding machine learning algorithms, the inductive approach is more practical as it avoids re-training the prediction algorithm for each test case, although it incurs a loss of statistical efficiency/power due to the splitting of the data \citep{toccaceli2019combination}.

\subsection{Local Coverage of Conformal Prediction}
Conformal prediction is mathematically guaranteed to produce prediction sets having finite-sample control over type I error rates, for any user-specified level of significance. However, while this property holds in aggregate over repeated sampling, this property does not necessarily hold conditionally on subsets of labels if $y$ is categorical, or conditionally on regions of values, if $y$ is real-valued \citep[see for instance][]{guan2023localized}. In particular, when the outcome $y$ suffers from severe class imbalance or is heavily skewed the SCP algorithm will tend to over-cover denser regions and under-cover sparser regions of $y$.  

One proposed method for achieving appropriate coverage on subsets of the label space is to use label-conditional conformal prediction, where the prediction set is calculated using only the non-conformity scores of $y$'s associated with one label at a time. This approach guarantees that the prediction set will have the correct coverage both in the aggregate and for each label of $y$ \citep{toccaceli2019combination}.  This has been shown to work well in practice, e.g., when predicting housing prices for different districts in a city \citep{hjort2023uncertainty}.

We extend the approach of label-conditional conformal prediction to real-valued $y$'s, where the labels can represent different regions in the $y$ space. In this case, the resulting prediction sets are calibrated to the local distribution of the calibration data in each label-class and will therefore, under the assumption of exchangeability, have the correct coverage in each label-class. 

\subsection{Bin-conditional Conformal Prediction}
Not all types of data have a natural clustering structure that can be used to condition the label-conditional conformal prediction algorithm. However, they may still suffer from non-uniform coverage of SCP sets across the range of the outcome space of $y$ \citep{guan2023localized}. This is especially a concern in cases where the outcome $y$ is real-valued and the distribution of $y$ is skewed, which is not uncommon in social science applications.

A field with notoriously skewed outcomes is the prediction of fatalities from armed conflict, where the number of fatalities is often zero but can also be very high in some cases \citep[see for instance][]{randahl2022inference,hegre2021views2020}. In this case, the prediction intervals may have too high coverage in the lower range of $y$ and too low coverage in the upper range of $y$. Additionally, the most salient predictions are often those that are in the upper range of $y$, and it is therefore important that the prediction intervals have the correct coverage in this range.

To solve the problem of non-uniform coverage across the range of $y$ in real-valued outcomes, we propose to extend the label-conditional conformal prediction approach to a BCCP approach, where the outcome space is partitioned into user-specified ranges (bins) on which the user wishes to obtain correct coverage. 

We specify the \textit{BCCP algorithm} for the inductive routine as follows:
\begin{enumerate}
    \item Partition the data into a training set and a calibration set.
    \item Train the prediction algorithm on the training set.
    \item Calculate the non-conformity score for each datum in the calibration set.
    \item Partition the non-conformity scores into user-specified bins based on the observed $y$'s in the calibration set.
    \item Set a user-specified coverage level $1-\alpha$ which to obtain.
    \item For each test case, run an SCP algorithm conditional on each bin separately, over a grid of hypothetical $y$ values associated with each bin.
    \item Calculate prediction interval\footnote{As this method is used for real-valued $y$, we refer to the resulting prediction sets as prediction intervals} for each bin by keeping the hypothetical $y$ values which produce non-conformity scores below the $1-\alpha$ quantile of the non-conformity scores of the $y$'s in the bin-conditional calibration set.
    \item Combine the prediction intervals from each bin to obtain the final prediction interval(s).

\end{enumerate}

This algorithm guarantees prediction intervals with correct coverage in each user-specified bin of $y$, assuming exchangeability of the data within each bin.

Two important questions arise when using the BCCP approach. The first is how to combine the prediction intervals from each bin to obtain the final prediction interval, and the second is how to partition the $y$'s into bins.

\subsubsection{Combining bins}
A potential drawback of the BCCP approach is that the final prediction set is not necessarily contiguous. Instead, this set may be a union of discontiguous prediction intervals. This is not necessarily a problem, but it may be less intuitive for the user or the decision-maker. We refer to this as the \textit{discontiguous prediction interval} method.

In the case where the prediction intervals are discontiguous, an alternative is to contiguize the intervals by using the lower and upper endpoints of the intervals across all bins. This approach guarantees that the prediction regions have \textit{at least} the correct coverage in each bin but may have higher coverage than specified by $\alpha$ in some regions of the outcome space and in the aggregate.

\subsubsection{Selecting the bins}

In the BCCP approach, the choice of bins is crucial, as it determines which ranges of the outcome space will receive well-calibrated prediction intervals. Bin regions may be completely arbitrarily set, including arbitrarily selected cut-off points, and do not need to be of equal size, neither in range nor in the number of observations per bin.\footnote{It would also be possible to define bins based on some other criteria than the outcome variable, for instance on some regions of covariates.} We argue that it is primarily the user's preferences and the specific application that should guide the choice of bins, as it is up to the user to decide which ranges of $y$ are of interest, although the user could also optimize the bin boundaries based on some criterion and cross-validation. While the bins can be defined arbitrarily, there are, however, some practical considerations that should be taken into account when defining bin boundaries.

First, the bins must be large enough to contain a sufficient number of data points to obtain reasonably precise estimates of the $(1 - \alpha)$-quantile of non-conformity scores. If a bin contains too few observations, the $(1 - \alpha)$-quantile of non-conformity scores used for calculating the prediction intervals may be imprecise, potentially leading to too wide prediction intervals.

Second, because the prediction intervals generated within the bins themselves may suffer from the same issue of non-uniform coverage as the SCP approach does on the aggregate level (BCCP with one bin is equivalent to SCP), the bins should be small enough to capture the local distribution of $y$ for the ranges of $y$ that the user is interested in. However, increasing the number of bins comes with a drawback: more bins will widen the prediction intervals because they enforce uniform coverage within each bin. This creates a fundamental trade-off: fewer bins lead to narrower intervals but may fail to provide consistent coverage across the outcome space, while more bins improve local calibration but result in wider intervals. The choice of bins should therefore balance the need for accurate local coverage with the goal of maintaining reasonably tight prediction intervals, considering the available data and the specific application. Furthermore, increasing the number of bins may also lead to a larger proportion of discontiguous prediction intervals, which may be less intuitive for the user when using the prediction intervals for decision-making. Alternatively, if the final intervals are contiguized, the higher number of bins will lead to a higher over-coverage rate in the aggregate.

\section{Simulation study}
To evaluate the performance of the BCCP algorithm, we conduct a simulation study to compare the coverage and width of the prediction intervals obtained from the BCCP algorithm to those obtained from the SCP algorithm across different values of $y$ in the test set.

We use a simple simulation setup with two predictor features $x_1$ and $x_2$ generated as independent uniform random variables between 0 and 1, and a prediction target $y$ which is generated as a log-normal distribution with parameters $\mu = x_1 + x_2$ and $\sigma = 0.5$ on the log-scale. The simulated distributions are heavily right-skewed with a large proportion of near-zero values.

We generate 10,000 instances of our variables and partition these into three sets: a training set with 5,000 instances, a calibration set with 2,500 instances, and a test set with 2,500 instances. We then train a simple linear regression model on each training set using the natural logarithm of $y$ as the target variable and $x_1$ and $x_2$ as the predictor variables. Predictions are made on the original $y$-scale by exponentiating the model's predictions.

Next, we compute the non-conformity scores on the calibration set, using the absolute prediction error as the non-conformity measure. We partition the calibration data into four bins based on the empirical 25th, 50th, and 75th percentiles of the $y$ values. To highlight the importance of the choice of bins we also consider two additional scenarios with two and six bins, again based on the empirical percentiles of the $y$ values in the calibration sets.\footnote{In practice, there is no strict requirement to use empirical percentiles for binning. Instead, the user can define bins based on regions of the outcome space where they expect the local distribution to differ, or which are of particular interest.}

We then apply the BCCP algorithm to the test set using both the contiguous (BCCPc) and discontiguous (BCCPd) interval approaches for $\alpha = 0.1$, giving an expected coverage rate of 90\%. The resulting prediction intervals are compared to those generated by the SCP algorithm, as well as to prediction intervals generated by alternative methods, including by bootstrapping errors from the calibration set (either on $y$ or $log(y)$ scale), quantile regression intervals, and parametric prediction intervals based on the log-normal distribution. For all methods, we evaluate both the overall coverage rate and the average width of the prediction intervals, as well as their performance across different ranges of $y$.

\subsection{Simulation results}
The mean aggregate and bin-conditional coverage of the simulation study are shown in Table \ref{tab:sim_results} below. The table displays both the aggregate coverage as well as the coverage in each of the four quartiles of $y$ for the SCP, BCCPc, and BCCPd algorithms. The results show that while the SCP algorithm achieves correct coverage in aggregate, it fails to obtain the correct coverage across different ranges of $y$. In contrast, the BCCP algorithms achieve at least the correct coverage in aggregate as well as within each bin it is conditional on.

The results of the BCCP algorithms with two bins are especially illustrative as they show that the coverage in both the first two and last two quartiles are correct in aggregate (i.e., marginally), but not within the individual quartiles. The BCCP algorithms with four and six bins, on the other hand, show that the coverage is at least at the nominal level for all quartiles as well as in aggregate. As expected, BCCP with contiguized intervals (BCCPc) slightly over-covers some of the quartiles of $y$, except when using only two bins the results are identical for the discontiguous and contiguous intervals. The discontiguous intervals (BCCPd) on the other hand, achieve coverage rates only marginally different from the nominal 90\% in all bins they are conditional on.

With regards to the alternative methods for generating prediction intervals, the results show that only the log-normal and quantile regression intervals achieve correct coverage in aggregate, while the other methods all produce coverage rates that are too low. That the parametric prediction intervals based on the log-normal distribution and quantile regression achieve the correct coverage in aggregate is not surprising, as the data was generated from a log-normal distribution. Strikingly, however, not even the log-normal or quantile regression prediction intervals achieve the correct coverage across the different ranges of $y$.

\begin{table}[H]
\centering
\caption{Coverage of the different prediction intervals in aggregate and across empirical quartiles (Q1-Q4) of $y$ in the simulation study. $\alpha=0.1$ and expected coverage is 0.90.}
\begin{tabular}{ll|cccc}
\hline
Method & Aggregate & Q1 of $y$ & Q2 of $y$ & Q3 of $y$ & Q4 of $y$ \\
\hline
  SCP & 0.9 & 0.99 & 0.98 & 0.99 & 0.64 \\
  BCCPc (2 bins) & 0.9 & 0.89 & 0.91 & 1.00 & 0.80 \\
  BCCPc (4 bins) & 0.91 & 0.90 & 0.90 & 0.95 & 0.90 \\
  BCCPc (6 bins) & 0.92 & 0.90 & 0.90 & 0.95 & 0.93 \\
  BCCPd (2 bins) & 0.9 & 0.89 & 0.91 & 1.00 & 0.80 \\
  BCCPd (4 bins) & 0.9 & 0.90 & 0.90 & 0.90 & 0.90 \\
  BCCPd (6 bins) & 0.9 & 0.89 & 0.90 & 0.91 & 0.88 \\
  Bootstrap & 0.84 & 1.00 & 1.00 & 0.95 & 0.42 \\
  Bootstrap (log) & 0.86 & 0.68 & 0.98 & 0.98 & 0.82 \\
  Log-normal & 0.9 & 0.82 & 0.98 & 0.98 & 0.82 \\
  Quantile Regression & 0.9 & 0.82 & 0.98 & 0.98 & 0.82 \\
\hline
\end{tabular}
\label{tab:sim_results}
\end{table}

\begin{figure}[H]
\centering
\includegraphics[width=0.8\textwidth]{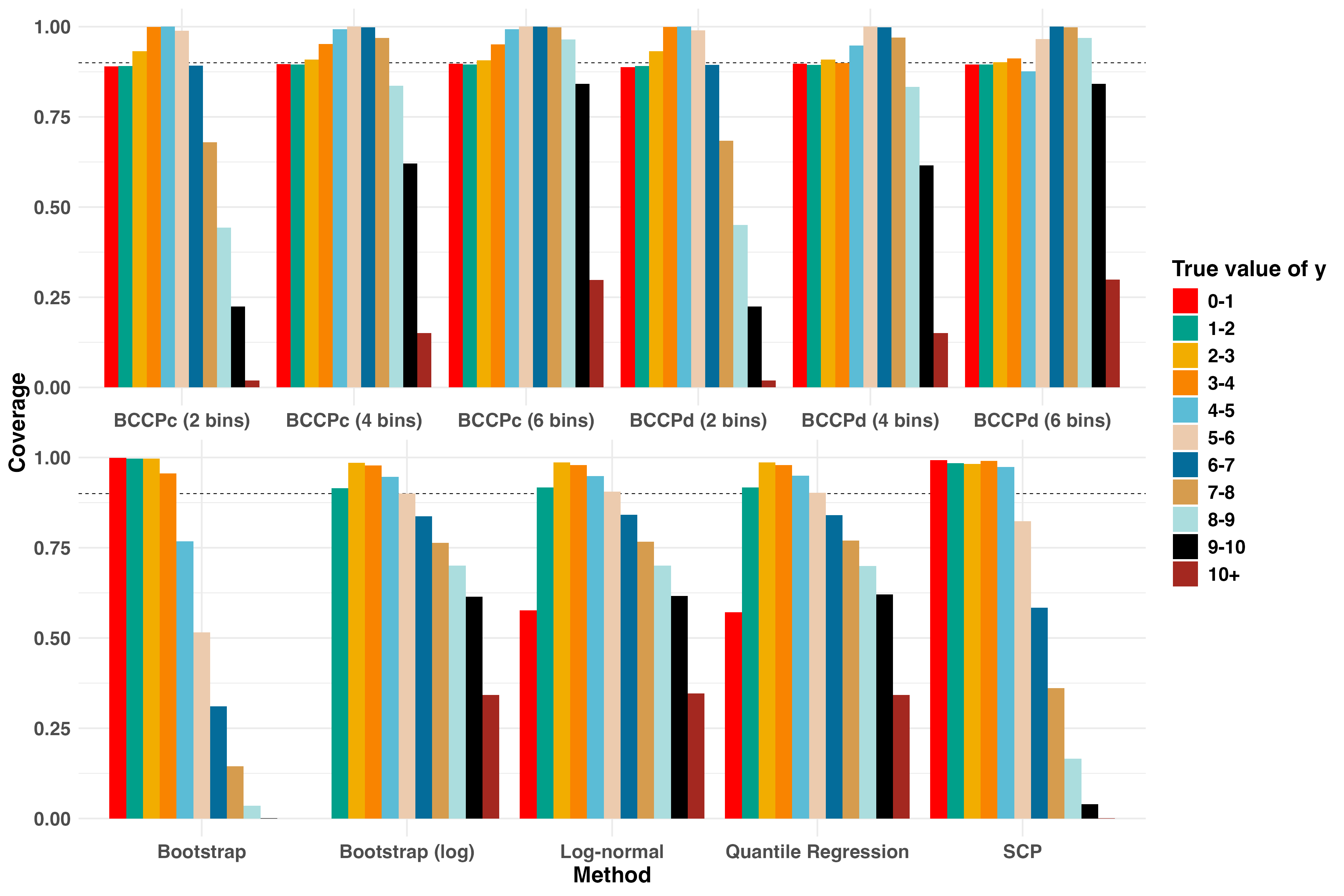}
\caption{Coverage across different values of $y$ for methods with correct aggregate coverage.}
\label{fig:sim_cov}
\end{figure}

The effect of bin selection on coverage can also be seen in Figure \ref{fig:sim_cov} below which shows the coverage rates across different values of the prediction target. The figure illustrates that as we increase the number of bins, the coverage becomes more uniform across the range of $y$, but that non-uniformity may still occur within the bins. For the highest values of $y$, the coverage is still much below the desired level even when using the 6 bin solution. The figure also shows that increasing the number of bins for the contiguized version of the BCCP algorithm (BCCPc) increases the over-coverage.

Looking at the SCP algorithm, we see that it severely over-covers values of $y$ near zero, where most of the data are located, and under-covers on the right tail of $y$, with the bootstrap method producing similar patterns of coverage. The bootstrap (log), quantile regression, and log-normal intervals produce near identical coverage patterns, with undercoverage at both tails and over-coverage in the middle of the distribution. The log-normal prediction intervals, which are clearly unsuited to the prediction problem, severly under-cover through the entire distribution.

\begin{figure}[H]
\centering
\includegraphics[width=0.8\textwidth]{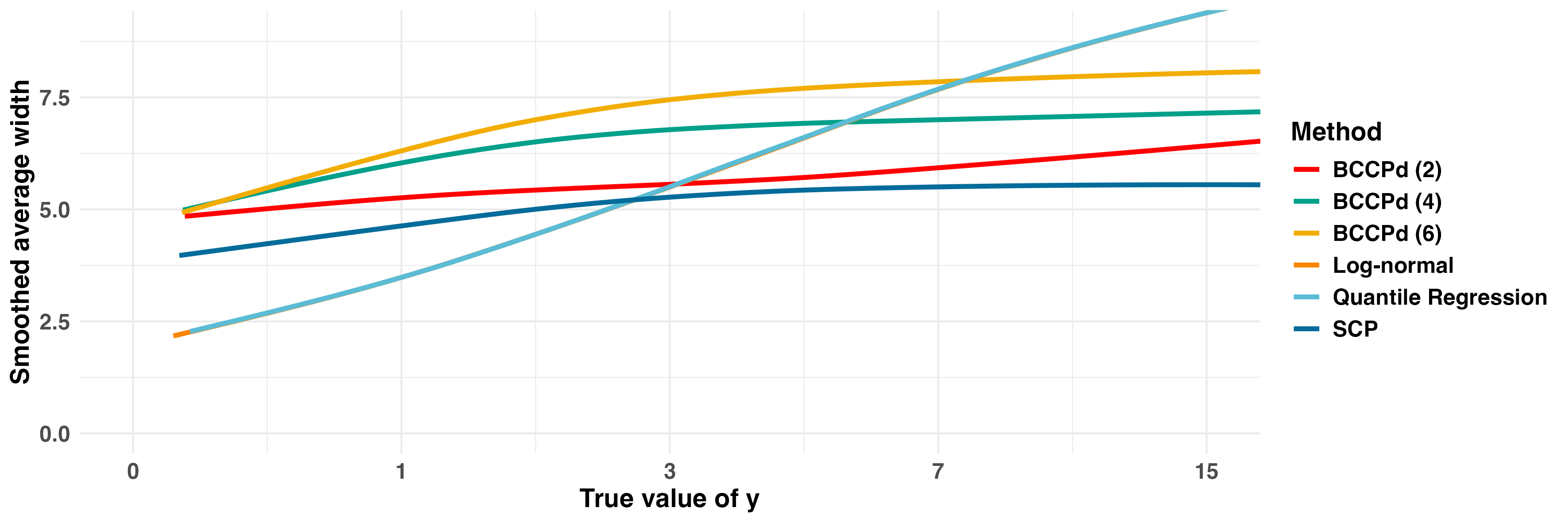}
\caption{Mean interval width across different values of $y$. Only the discontiguous version of the BCCP algorithm is shown for legibility. $\alpha=0.1$ and expected coverage is 0.90.}
\label{fig:sim_width}
\end{figure}

The trade-off between coverage and width is illustrated in Figure \ref{fig:sim_width} which shows the smoothed mean width of the prediction intervals across the range of predicted $y$ for the prediction interval methods with correct aggregate coverage. The figure shows that the SCP algorithm produces the narrowest intervals among the conformal methods, and that increasing the number of bins in the BCCP method also increases the average width of the intervals. This demonstrates the tradeoff between maintaining appropriate coverage and the width of the intervals. The quantile regression and log-normal intervals are, as expected, near identical, and are narrower for lower values of $y$ but increase in width faster than the conformal counterparts.

\section{BCCP of fatalities from armed conflict}
To test the practical application of the BCCP algorithm and compare it to the SCP algorithm, we apply both methods to a prediction model for fatalities from armed conflict. Specifically, we apply the algorithms to one of the constituent models from the Violence Early Warning System (VIEWS) fatalities forecasting project \citep{Hegre2022FCDO}. The model we use is the broad topics constituent model, inspired by \citet{mueller2018reading}, which consists of a set of 64 predictor features including a number derived from 15 different topics covered in news articles. The prediction target is the number of fatalities from armed conflict in which at least one of the actors is a state \citep{davies2023organized}, for each country month between the years 2000 and 2022.

The prediction target in this application exhibits characteristics that may be challenging from a prediction standpoint. In particular, the prediction target is severely zero-inflated and highly right-skewed with approximately 86.7\% being zeros and the top 1\% of the observations accounting for nearly 75\% of the fatalities.

We obtain the predictions using a standard random forest regressor. In line with VIEWS standards, we make predictions for the number of fatalities on the log1p scale, i.e., log(fatalities+1) \citep{Hegre2022FCDO}. To test whether the prediction intervals we obtain have a correct level of coverage, we run a set of 1,000 simulations where we randomly partition our data into training (70\%), calibration (20\%), and test (10\%).\footnote{In these simulations we assume that predictions from the models are \textit{exchangeable}; we thus ignore the panel structure of the data. We recognize that this is a strong assumption which is unlikely to hold fully in practice. However, the purpose of this exercise is to test the performance of the prediction intervals in a controlled setting. We also tested the BCCP algorithm on a standard (single) time series split of the data and found that the results were similar to the results presented here in that the BCCP algorithm produced prediction intervals with the near correct level of coverage, albeit with a slight under-coverage in the tails, while the SCP and other alternatives severely over-covered zeros and under-covered the right tail. The results of this test can be made available on request. We discuss the plausibility of the exchangeability assumption and possible extensions to deal with this problem in the discussion section.}

In each simulation, we train the random forest model using the randomly assigned training data, we then calculate the non-conformity scores by making out-of-sample prediction on the calibration data, and finally we calculate the prediction intervals on the test data and compute the coverage of the prediction intervals. We run both an SCP algorithm and the discontiguous BCCP algorithm using 2, 4, and 7 bins\footnote{We use the discontiguous version of the BCCP algorithm as it is the more principled option which is expected to yield the nominal level of coverage rather than a slight over-coverage. Tests of the contiguized BCCP algorithm showed that the performances of the methods do not differ substantially. The results for the contiguized version of the BCCP algorithm can be made available on request.}, as well as alternative algorithms for obtaining the prediction intervals, including quantile forest intervals, bootstrapped intervals, and different parametric intervals.

In the two-bin version, we split the calibration data into bins based on true zeros versus non-zeros. In the seven-bin versions we keep one bin for zeros and then split the non-zeros into six bins based on an exponentially increasing number of fatalities in each bin such that the bins are defined in the following way; bin 1: 0 fatalities, bin 2: 1--2 fatalities, bin 3: 3--7 fatalities, bin 4: 8--20 fatalities, bin 5: 21--54 fatalities, bin 6: 55--148 fatalities, and bin 7: 149+ fatalities. In the four-bin version, we use the same binning structure as the seven-bin version but combine bins 2--3, 4--5, and 6--7. The binning structure we have chosen is essentially arbitrary, but does result in bins with approximately similar number of observations outside the bin for zeros. We do not claim that this is the optimal binning structure for these data but rather a first attempt to explore the potential of the BCCP algorithm on this type of data. Across all simulations we use $\alpha = 0.1$ giving an expected coverage of 90\% for the prediction intervals.

A further issue with this prediction problem is that since the prediction target is the log1p transformed counts of fatalities, the prediction target is not strictly continuous but rather quasi-continuous, where the true values are restricted to the log1p transformation of all positive integers. This also poses a challenge to the continuous conformal prediction algorithm since it limits the valid values of the prediction interval. To deal with this problem, we make our predictions and prediction intervals on the original fatality scale by reversing the log1p transformation\footnote{i.e., $e^{\hat{y}}-1$.} for the predictions and the prediction intervals. To ensure that the prediction intervals correspond to this data structure, we round the lower and upper bounds of the prediction intervals to the nearest integer.\footnote{This approach may lead to a slight under- or over-coverage of the prediction intervals, but we believe that this is a reasonable approximation given the quasi-continuous nature of the prediction target.}

\subsection{Results}
We begin by comparing the coverage of the algorithms in aggregate as well as across true zeros and non-zeros. These results can be seen in Table \ref{tab:raw} below. These results exhibit the same pattern as in the simulation study, i.e., that while the SCP algorithm produces an aggregate coverage level close to the nominal level\footnote{The slight over-coverage for the SCP algorithm in the aggregate is likely related to the rounding of the interval edges.}, it both severely over-covers the zeros and under-covers the non-zeros. The BCCP algorithms, on the other hand, manage to produce an appropriate level of coverage among both zeros and non-zeros.

For the other methods, the results are mixed. The bootstrapping and parametric methods all produce coverage levels which are near the user-specified level in aggregate, but severely over-cover the zeros and under-cover the non-zero observations. The quantile regression intervals cover all zeros, and approximately the correct proportion of non-zero observations. Interestingly, the bootstrapped (log) intervals have nearly identical coverage rates as the SCP algorithm.

\begin{table}[!htbp] \centering
  \caption{Coverage of prediction intervals. $\alpha=0.1$ and expected coverage is 0.90.}
  \label{tab:raw}
\begin{tabular}{ll|l|ll}
\hline
Method & Aggregate & Zeros & Non-zeros\\
  SCP & 0.92 & 0.98 & 0.52 \\
BCCPd (2 bins) & 0.90 & 0.90 & 0.90 \\
  BCCPd (4 bins) & 0.90 & 0.90 & 0.91 \\
  BCCPd (7 bins) & 0.90 & 0.90 & 0.90 \\
  Bootstrap (log) & 0.92 & 0.98 & 0.51 \\
  Bootstrap & 0.90 & 1.00 & 0.29 \\
  Log-normal & 0.94 & 0.99 & 0.63 \\
  Negative Binomial & 0.92 & 0.99 & 0.44 \\
  Poisson & 0.90 & 0.99 & 0.32 \\
  Quantreg & 0.99 & 1.00 & 0.89 \\
   \hline
\hline
\end{tabular}
\end{table}

Further dividing the results into the value ranges corresponding to the seven bins used for the BCCP algorithm, we again see that the coverage remains appropriate for the number of bins used. These results are shown in Figure \ref{fig:bin} below. This further highlights the flexibility of the BCCP algorithm and the importance of setting the number of bins so that the coverage is appropriate across the most important ranges of the prediction target. For the alternative methods, we see that the quantile regression intervals maintain an appropriate level of coverage across all bins except the first, where it covers all observations, and the last, where it covers only 80\% of the observations. The other methods all exhibit the same pattern with over-coverages of zeros and ever increasing under-coverages of non-zeros as the number of fatalities increases.

\begin{figure}[H]

\includegraphics[width=\textwidth]{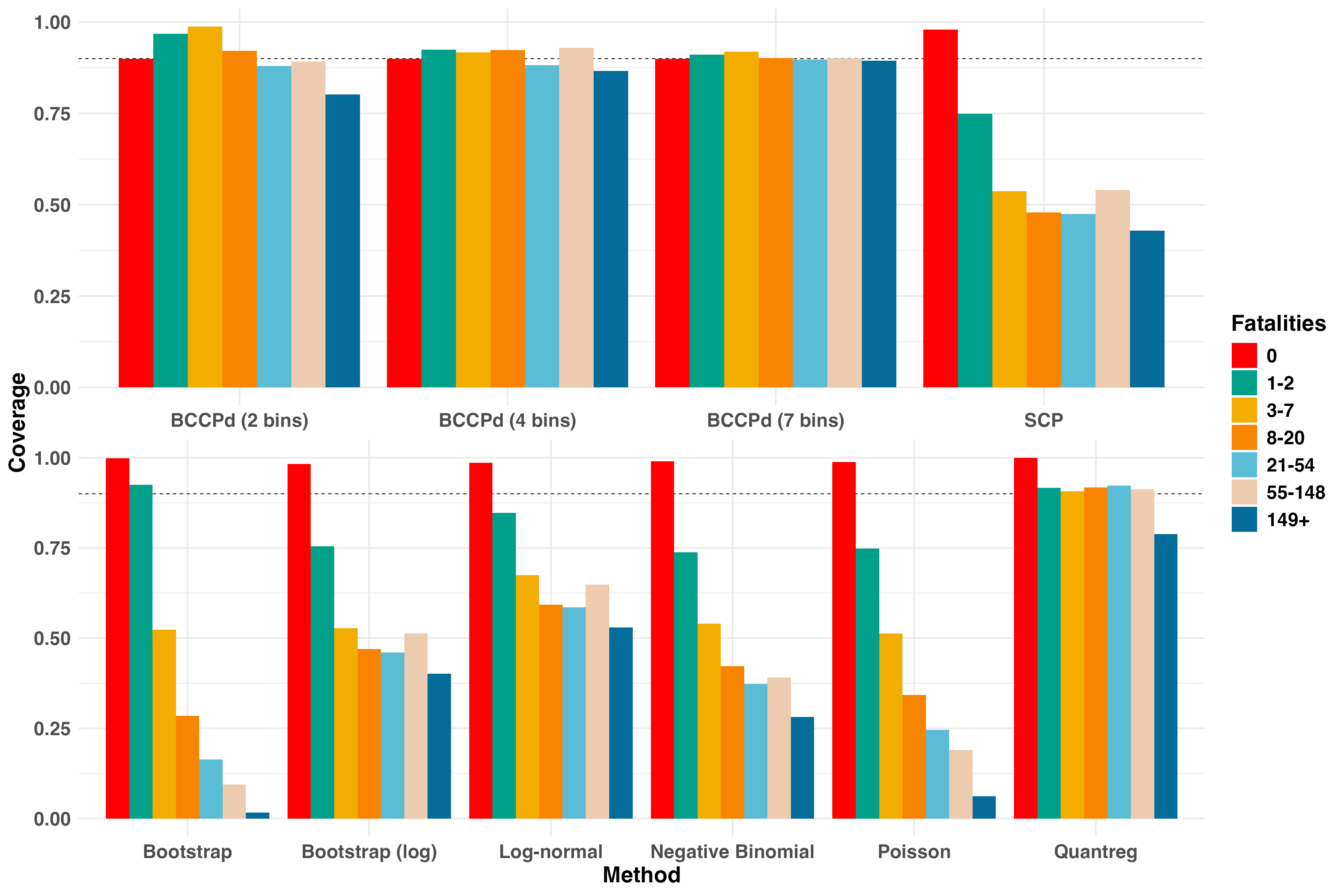}

\caption{Coverage of prediction intervals across the seven bins in the data. Only the discontiguous versions of the BCCP algorithm are shown.}
\label{fig:bin}
\end{figure}

Lastly, looking at the average width of the prediction intervals, we see that the BCCP algorithm produces wider intervals than the alternative methods. This is shown in Figure \ref{fig:width} below. This result is in line with the simulation study, where we saw that the BCCP algorithm produced wider intervals than the SCP algorithm, and that increasing the number of bins also increased the width of the intervals. This is a tradeoff where the user needs to weigh the uniformity of the coverage over various regions with the width of the resulting intervals.

\begin{figure}[H]
\includegraphics[width=\textwidth]{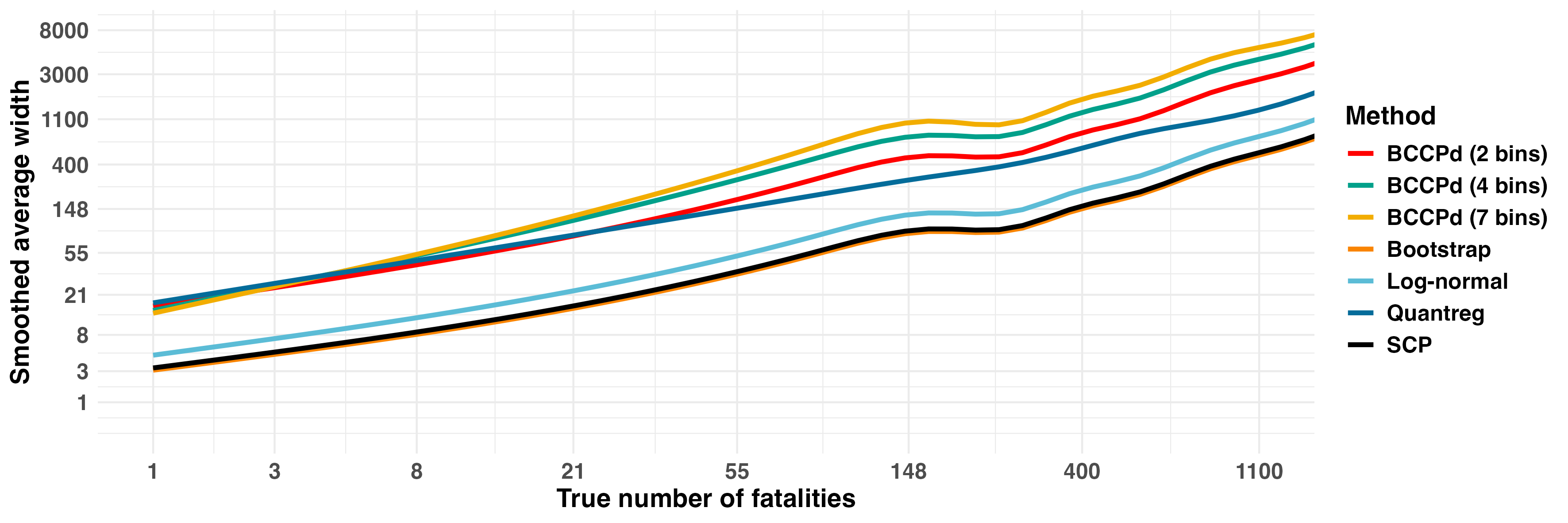}

\caption{Mean width of prediction intervals across different number of fatalities.}
\label{fig:width}
\end{figure}

\section{Discussion}
In this paper, we have introduced a new conformal prediction algorithm that maintains an appropriate level of coverage across different ranges of the prediction target. When the prediction target has a heavily right-skewed and/or zero-inflated distribution, SCP approaches fail in producing correct coverage uniformly over the outcome space. We demonstrated that our \textit{BCCP} method achieves correct coverage within user-defined bins of values, although at the cost of wider prediction intervals. The method performed well in both a simulation study and in a real-world application predicting the number of fatalities from armed conflict.

Furthermore, this paper has highlighted the need for developing methods that allow researchers to produce forecasts which not only give point predictions, but also accurate and interpretable prediction intervals. The BCCP algorithm extends the applicability of conformal prediction algorithms to a wider range of prediction problems, as it allows the user to define the most important ranges of the prediction target and ensures that the prediction intervals are calibrated within these ranges. This may be particularly important in fields where the end users of the predictions are not experts in statistics and machine learning but rather stakeholders or policymakers who may need to make decisions based on the predictions. In this paper we have shown the utility of this method for right-skewed count data (fatalities) but this framework can be used regardless of many distributional characteristics of the outcome, including for estimating uncertainty in predicted probabilities.

The algorithms within the conformal prediction framework assume that the predictions from the models are \textit{exchangeable}, meaning that their order does not matter. In the context of conflict prediction, this assumption may not hold as the data generating process may change over time. Despite this potential limitation, the strong performance of the BCCP algorithm in the real-world application suggests that the method may be robust to some violations of this assumption. Additionally, the general approach presented in this paper may also easily be combined with other methods developed for conformal prediction in applications which do not necessarily fulfill the exchangeability assumption \citep[see for instance][]{barber2023conformal,mao2024valid,hjort2023uncertainty}.

Selecting the number of bins to use in the BCCP algorithm is a crucial step in the application of the method. In this paper, we have argued that the number of bins should be set by the researcher based on the data and research question. Essentially, the method requires the bins to be large enough to contain a sufficient number of observations for the $(\alpha/2, 1-\alpha/2)$ quantiles of non-conformity scores to be estimated with reasonable accuracy in the calibration data, but beyond that the number of bins and their cut-points can be arbitrarily set by the researcher. While it is possible to simply set a number of observations to be achieved per bin and then maximize the number of bins, we caution against this approach as the number of bins will affect the width of the prediction intervals. Instead, the number of bins should be set such that an appropriate level of coverage is achieved for the most important ranges of the prediction target, defined by the researcher. However, given a specific application and evaluation metric, it may also be possible to optimize the number of bins in the BCCP algorithm by treating it, and the cut-points, as a hyperparameters. Future research should investigate methods for selecting the number of bins in the BCCP algorithm in such a data-driven manner, for instance by using cross-validation or other model selection techniques.

We have made the BCCP algorithm available in the \texttt{pintervals} package for R on CRAN.

\newpage

\printbibliography

@article{hjort2023uncertainty,
  title={Uncertainty quantification in automated valuation models with locally weighted conformal prediction},
  author={Hjort, Anders and Hermansen, Gudmund Horn and Pensar, Johan and Williams, Jonathan P},
  journal={arXiv preprint arXiv:2312.06531},
  year={2023}
}

@article{zhang2020random,
  title={Random forest prediction intervals},
  author={Zhang, Haozhe and Zimmerman, Joshua and Nettleton, Dan and Nordman, Daniel J},
  journal={The American Statistician},
  year={2020},
  publisher={Taylor \& Francis}
}

@article{hesterberg2015teachers,
  title={What teachers should know about the bootstrap: Resampling in the undergraduate statistics curriculum},
  author={Hesterberg, Tim C},
  journal={The American Statistician},
  volume={69},
  number={4},
  pages={371--386},
  year={2015},
  publisher={Taylor \& Francis}
}

@article{tian2022methods,
  title={Methods to compute prediction intervals: A review and new results},
  author={Tian, Qinglong and Nordman, Daniel J and Meeker, William Q},
  journal={Statistical Science},
  volume={37},
  number={4},
  pages={580--597},
  year={2022},
  publisher={Institute of Mathematical Statistics}
}

@article{hegre2019views,
  title={ViEWS: A political violence early-warning system},
  author={Hegre, Håvard and Allansson, Marie and Basedau, Matthias and Colaresi, Michael and Croicu, Mihai and Fjelde, Hanne and Hoyles, Frederick and Hultman, Lisa and Högbladh, Stina and Jansen, Remco and others},
  journal={Journal of peace research},
  volume={56},
  number={2},
  pages={155--174},
  year={2019}
}

@article{hegre20242023,
  title={The 2023/24 VIEWS Prediction Challenge: Predicting the Number of Fatalities in Armed Conflict, with Uncertainty},
  author={Hegre, H{\aa}vard and Vesco, Paola and Colaresi, Michael and Vestby, Jonas and Timlick, Alexa and Kazmi, Noorain Syed and Becker, Friederike and Binetti, Marco and Bodentien, Tobias and Bohne, Tobias and others},
  journal={arXiv preprint arXiv:2407.11045},
  year={2024}
}

@Misc{Hegre2022FCDO,
  author       = {Hegre, Håvard and Dale, James and Randahl, David and Croicu, Mihai and Jansen, Remco and Vesco, Paola and Lindqvist-McGowan, Angelica and Landsverk, Peder and Leis, Maxine and Mueller, Hannes and Rauh, Christopher and Geelmuyden Rod, Espen and Akbari, Forogh and Gåsste, Tim and Rakhmankulova, Malika},
  date         = {2022},
  title        = {Forecasting Fatalities},
  howpublished = {Uppsala: working paper},
  url          = {https://www.diva-portal.org/smash/get/diva2:1667048/FULLTEXT01.pdf}
}

@article{mao2024valid,
  title={Valid model-free spatial prediction},
  author={Mao, Huiying and Martin, Ryan and Reich, Brian J},
  journal={Journal of the American Statistical Association},
  volume={119},
  number={546},
  pages={904--914},
  year={2024},
  publisher={Taylor \& Francis}
}

@article{barber2023conformal,
  title={Conformal prediction beyond exchangeability},
  author={Barber, Rina Foygel and Candes, Emmanuel J and Ramdas, Aaditya and Tibshirani, Ryan J},
  journal={Annals of Statistics},
  volume={51},
  number={2},
  pages={816--845},
  year={2023},
  publisher={Institute of Mathematical Statistics}
}

@article{guan2023localized,
  title={Localized conformal prediction: A generalized inference framework for conformal prediction},
  author={Guan, Leying},
  journal={Biometrika},
  volume={110},
  number={1},
  pages={33--50},
  year={2023},
  publisher={Oxford University Press}
}

@article{fontana2023conformal,
  title={Conformal prediction: a unified review of theory and new challenges},
  author={Fontana, Matteo and Zeni, Gianluca and Vantini, Simone},
  journal={Bernoulli},
  volume={29},
  number={1},
  pages={1--23},
  year={2023},
  publisher={Bernoulli Society for Mathematical Statistics and Probability}
}

@inproceedings{vovk1999machine,
  title={Machine-Learning Applications of Algorithmic Randomness},
  author={Vovk, Volodya and Gammerman, Alexander and Saunders, Craig},
  booktitle={Proceedings of the Sixteenth International Conference on Machine Learning},
  pages={444--453},
  year={1999}
}

@inproceedings{saunders1999transduction,
  title={Transduction with confidence and credibility},
  author={Saunders, C and Gammerman, A and Vovk, V},
  booktitle={Proceedings of the 16th international joint conference on Artificial intelligence-Volume 2},
  pages={722--726},
  year={1999}
}

@article{shafer2008tutorial,
  title={A Tutorial on Conformal Prediction.},
  author={Shafer, Glenn and Vovk, Vladimir},
  journal={Journal of Machine Learning Research},
  volume={9},
  number={3},
  year={2008}
}

@book{vovk2005algorithmic,
  title={Algorithmic learning in a random world},
  author={Vovk, Vladimir and Gammerman, Alexander and Shafer, Glenn},
  year={2022},
  edition={2},
  publisher={Springer}
}

@article{williams2024,
  title={Bayesian hidden {M}arkov models for latent variable labeling assignments in conflict research: {A}pplication to the role ceasefires play in conflict dynamics},
  author={Williams, Jonathan P and Hermansen, Gudmund H and Strand, H{\aa}vard and Clayton, Govinda and Nyg{\aa}rd, H{\aa}vard Mokleiv},
  journal={Annals of Applied Statistics},
  volume={18},
  number={3},
  pages={2034--2061},
  year={2024},
  publisher={Institute of Mathematical Statistics}
}

@article{dey2023conformal,
  title={Conformal Prediction for Text Infilling and Part-of-Speech Prediction},
  author={Dey, Neil and Ding, Jing and Ferrell, Jack and Kapper, Carolina and Lovig, Maxwell and Planchon, Emiliano and Williams, Jonathan P},
  journal={The New England Journal of Statistics in Data Science},
  volume={1},
  pages={69--83},
  year={2023},
  publisher={New England Statistical Society}
}

@article{toccaceli2019combination,
  title={Combination of inductive {M}ondrian conformal predictors},
  author={Toccaceli, Paolo and Gammerman, Alexander},
  journal={Machine Learning},
  volume={108},
  pages={489--510},
  year={2019},
  publisher={Springer}
}

@incollection{papadopoulos2008inductive,
  title={Inductive conformal prediction: Theory and application to neural networks},
  author={Papadopoulos, Harris},
  booktitle={Tools in artificial intelligence},
  year={2008},
  publisher={Citeseer}
}

@article{mueller2018reading,
  title={Reading between the lines: Prediction of political violence using newspaper text},
  author={Mueller, Hannes and Rauh, Christopher},
  journal={American Political Science Review},
  volume={112},
  number={2},
  pages={358--375},
  year={2018},
  publisher={Cambridge University Press}
}

@article{chernozhukov2021distributional,
  title={Distributional conformal prediction},
  author={Chernozhukov, Victor and W{\"u}thrich, Kaspar and Zhu, Yinchu},
  journal={Proceedings of the National Academy of Sciences},
  volume={118},
  number={48},
  pages={e2107794118},
  year={2021},
  publisher={National Academy of Sciences}
}

@article{sesia2021conformal,
  title={Conformal prediction using conditional histograms},
  author={Sesia, Matteo and Romano, Yaniv},
  journal={Advances in Neural Information Processing Systems},
  volume={34},
  pages={6304--6315},
  year={2021}
}

@article{randahl2022inference,
  title={Inference with extremes: Accounting for Extreme Values in Count Regression Models},
  author={Randahl, David and Vegelius, Johan},
  journal={International Studies Quarterly},
  year={forthcoming}
}

@article{davies2023organized,
  title={Organized violence 1989--2022, and the return of conflict between states},
  author={Davies, Shawn and Pettersson, Ther{\'e}se and {\"O}berg, Magnus},
  journal={Journal of Peace Research},
  pages={00223433231185169},
  year={2023},
  publisher={SAGE Publications Sage UK: London, England}
}

@article{hegre2021views2020,
  title={ViEWS2020: revising and evaluating the ViEWS political violence early-warning system},
  author={Hegre, H{\aa}vard and Bell, Curtis and Colaresi, Michael and Croicu, Mihai and Hoyles, Frederick and Jansen, Remco and Leis, Maxine Ria and Lindqvist-McGowan, Angelica and Randahl, David and R{\o}d, Espen Geelmuyden and others},
  journal={Journal of peace research},
  volume={58},
  number={3},
  pages={599--611},
  year={2021},
  publisher={SAGE Publications Sage UK: London, England}
}

@article{vesco2022united,
  title={United they stand: Findings from an escalation prediction competition},
  author={Vesco, Paola and Hegre, H{\aa}vard and Colaresi, Michael and Jansen, Remco Bastiaan and Lo, Adeline and Reisch, Gregor and Weidmann, Nils B},
  journal={International Interactions},
  volume={48},
  number={4},
  pages={860--896},
  year={2022},
  publisher={Taylor \& Francis}
}

@article{morgan2019varieties,
  title={Varieties of forecasts: Predicting adverse regime transitions},
  author={Morgan, Richard and Beger, Andreas and Glynn, Adam},
  journal={V-Dem Working Paper},
  volume={89},
  year={2019}
}

@article{bell2016coup,
  title={Coup d’{\'e}tat and democracy},
  author={Bell, Curtis},
  journal={Comparative Political Studies},
  volume={49},
  number={9},
  pages={1167--1200},
  year={2016},
  publisher={Sage Publications Sage CA: Los Angeles, CA}
}
\end{document}